\documentclass[reprint,twocolumn,superscriptaddress, showpacs,aps,pra]{revtex4-1}
\usepackage{hyperref}
\usepackage[T1]{fontenc}
\usepackage{amssymb}
\usepackage{amsmath}
\usepackage{graphicx}
\usepackage{xcolor}
\usepackage{bbm}

\newcommand{\unity}{\mathbbm{1}}
\newcommand{\bra}[1]{\left \langle #1 \right |}
\newcommand{\ket}[1]{\left | #1 \right \rangle}
\newcommand{\braket}[2]{\left \langle #1  | #2 \right \rangle}

\begin{document}
\title{Short time behavior of continuous time quantum walks on graphs}

\date{\today}
\author{Bal\'azs Endre Szigeti}
\affiliation{Institute for Particle and Nuclear Physics, Wigner Research Centre for Physics, Konkoly-Thege Mikl\'{o}s \'{u}t 29-33, H-1121 Budapest, Hungary}
\affiliation{Department of Atomic Physics, E\" otv\" os
University, P\'azm\'any P\'eter S\'et\'any 1/A, H-1117 Budapest, Hungary}
\author{G\'abor Homa}
\affiliation{Department of Physics of Complex Systems, E\" otv\" os
University, P\'azm\'any P\'eter S\'et\'any 1/A, H-1117 Budapest, Hungary}
\author{Zolt\'{a}n Zimbor\'{a}s}
\affiliation{Institute for Particle and Nuclear Physics, Wigner Research Centre for Physics, Konkoly-Thege Mikl\'{o}s \'{u}t 29-33, H-1121 Budapest, Hungary}
\affiliation{MTA-BME Lend\"ulet Quantum Information Theory Research Group}
\affiliation{Mathematical Institute, Budapest University of Technology and Economics, \\
P.O.Box 91, H-1111, Budapest, Hungary.}
\author{Norbert Barankai}
\affiliation{MTA-ELTE Theoretical Physics Research Group, P\'azm\'any P\'eter S\'et\'any 1/A, H-1117 Budapest, Hungary}

\begin{abstract}
Dynamical evolution of systems with sparse Hamiltonians can always be recognized as continuous time quantum walks (CTQWs) on  graphs. In this paper, we analyze the short time asymptotics of CTQWs. In recent studies, it was shown that for the classical diffusion process the short time asymptotics of the transition probabilities follows power laws whose exponents are given by the usual combinatorial distances of the nodes. Inspired by this result, we perform a similar analysis for CTQWs both in closed and open systems, including time-dependent couplings. For time-reversal symmetric coherent quantum evolutions, the short time asymptotics of the transition probabilities is completely determined by the topology of the underlying graph analogously to the classical case, but with a doubled power-law exponent. Moreover, this result is robust against the introduction of on-site potential terms. However, we show that time-reversal symmetry breaking terms and non-coherent effects can significantly alter the short time asymptotics. The analytical formulas are checked against numerics, and excellent agreement is found. Furthermore, we discuss in detail the relevance of our results for quantum evolutions on particular network topologies.
\end{abstract}
\maketitle

\section{Introduction}
Continuous time quantum walks (CTQWs) on graphs \cite{kempe2003quantum, mulken2011continuous, venegas12quantum, reitzner2011quantum} have been used frequently in the past to successfully model coherent transport phenomena in those systems whose phenomenological description allows the application of tight-binding approximations \cite{may2008charge}. Examples of such exciton networks consists of light harvesting complexes \cite{mohseni2008environment, caruso2009highly}, dendrimers \cite{mulken2006coherent}, trapped atomic ions \cite{PhysRevLett.122.050501} and arrays of quantum dots \cite{perdomo2010engineering, semiao2010vibration}, just to name a few.  

From a quantum information perspective, CTQWs appeared as possible physically realizable implementations of  quantum algorithms of search \cite{farhi1998quantum, childs2004spatial, portugal2013quantum, meyer2015connectivity, PhysRevLett.116.100501} and generic quantum computation \cite{PhysRevLett.102.180501,chase2008universal,agliari2010quantum}, and were compared on various occasions with their classical counterpart, the continuous time random walk (CTRW), that is, the diffusion process \cite{chung1997spectral, weiss2005aspects, telcs2006art}. 

A large number of experiments \cite{engel2007evidence,collini2010coherently,panitchayangkoon2010long,govia2017quantum}, numerical calculations and theoretical studies \cite{mulken2011continuous, wang2013physical, kendon2007decoherence, zimboras2013quantum, faccin2013degree, rossi2017continuous, boada2017quantum} have been devoted to analyze the transport properties of these systems. Among the most investigated topics were the {\it state transfer properties} \cite{bose2003quantum, christandl2004perfect, burgarth2005conclusive, bernasconi2008quantum,cameron2014universal, liu2015quantum} and the {\it long time behavior} \cite{konno2005limit, shikano2013, darazs2013time, philipp2016continuous, Fedichkin2005, liu2017steady} of these systems. Closed as well as open systems were studied and now there are plenty of examples where the supremacy of  CTQW over CTRW has been demonstrated. However, there are some cases when CTQWs underperform the old diffusive transport \cite{xu2009exact,agliari2008dynamics}.

Contrary to the long time asymptotics, the behavior of CTQWs at short timescales has missed such substantial attention. This is especially surprising if one notes that the short time dynamics of local Hamiltonians appearing in universal, continuous time quantum computation offers non-tomographical, efficient reconstruction of the governing Hamiltonian \cite{shabani2011estimation,da2011practical}. This resembles the situation in the theory of CTRW: Though the study of the short time asymptotics of Brownian motion on Riemannian manifolds has been initiated nearly half a century ago \cite{varadhan1967diffusion} and the results obtained have been subsequently extended and generalized in many ways \cite{molchanov1975diffusion,10.2307/2652939,terElst2007}, theorems concerning short time behavior of CTRW on graphs have been appeared only recently. In two current studies \cite{keller2016note,2018arXiv180102183S}, it was shown that the short time behavior of the transition probabilities of diffusion processes differ in a considerable amount when compared to their (in space) continuous  counterpart. While Brownian motion in locally Euclidean spaces can be approximated by a Gaussian distribution for short timescales, the same type of asymptotics tells that the transition probabilities $p(y,t|x)$, corresponding to distinct vertices $x$ and $y$ of a graph follow power law. If $d(x,y)$ is the distance between the aforementioned vertices, then \cite{keller2016note,2018arXiv180102183S}
\begin{equation}
\lim_{t\rightarrow 0}\frac{\ln p(y,t|x)}{\ln t}=d(x,y),
\end{equation} 
i.e., for small positive times $t$ we have 
\begin{equation}
p(y,t|x)=c(x,y)t^{d(x,y)}+\mathcal{O}(t^{d(x,y)+1}).
\label{EQ1}
\end{equation} 
Also the constant $c(x,y)$ has been determined \cite{2018arXiv180102183S}. If $\ell(x,y)$ denotes the number of shortest paths that connects $x$ to $y$, then 
\begin{equation}
c(x,y)=\frac{\ell(x,y)}{d(x,y)!}.
\end{equation}

In this paper, we show that similar results apply to CTQWs as well. Given a tight-binding model with adjacency matrix $A$ and on-site potential $V$, the complex transition amplitudes of the CTQW between position eigenstates $\ket{x}$ and $\ket{y}$ follows the asymptotics
\begin{equation}
\bra{x}U(t)\ket{y}=\frac{\ell(x,y)}{d(x,y)!}(-\mathrm{i}t)^{d(x,y)}+\mathcal{O}(t^{d(x,y)+1}).
\label{EQ3}
\end{equation}  
Thus, the time evolution of the entries of the mixing matrix $M_{xy}(t)=|\bra{x}U(t)\ket{y}|^2$ of the CTQW possesses the short time asymptotic form
\begin{equation}
M_{xy}(t)=\left(\frac{\ell(x,y)}{d(x,y)!}\right)^2t^{2d(x,y)}+\mathcal{O}(t^{2d(x,y)+1}).
\label{EQ2}
\end{equation}  
Since $M_{xy}(t)$ is the probability of finding the system in the position eigenstate $\ket{y}$ if initially it was prepared in the state $\ket{x}$, the comparison of Eq.~\eqref{EQ1} and Eq.~\eqref{EQ2} shows that CTQWs always underperform CTRWs at short timescales. Such a douling effect has been also observed in the tail distribution of the first passage time of CTQW \cite{thiel2018first}: The long-time assymptotics of the first passage time of a quantum walker of a one dimensional tight-binding model follows a power law in time with exponent $-3$, while a classic result of Lévy shows that such a scaling in CTRW has exponent $-3/2$ \cite{thiel2018spectral}. This is a rather general phenomenon which can appear when the spectrum of the Hamiltionian is continuous and the so called measurement density of states contains Van Hove singularities \cite{thiel2018spectral}. The sort-time analysis of the evolution of CTQWs coupled to its environment with the assumption of Markovian open system dynamics shows that a small amount of decoherence can halve the exponent in Eq.~\eqref{EQ2} to that of Eq.~\eqref{EQ1}, resembling the well studied properties of environment assisted quantum transport \cite{shabani2011estimation,da2011practical}.

Since the set of Hamiltonians is much larger than the set of symmetric generators of stochastic Markovian dynamics, the structure of the short time asymptotics of CTQWs is more abundant compared to that of CTRWs. These noticeable differences, caused by interference patterns become apparent when one considers chiral quantum walks \cite{zimboras2013quantum,lu2016chiral}. It turns out that the interference patterns can increase the exponent in Eq.~\eqref{EQ2} resulting in further deceleration of the initial dynamics.

Interestingly, the asymptotics of Eq.~\eqref{EQ3} is universal in the sense that the coefficients appearing do not depend on the on-site potential. The potential matrix $V$ only determines the timescale of the short-time regime, where Eq.~\eqref{EQ3} is worth to consider. Note, however, that a closer look on the evolution and the application of time-dependent perturbation theory can further improve Eq.~\eqref{EQ3} and widen the time horizon where results like Eq.~\eqref{EQ3} can approximate the initial dynamics.

The paper is organized as follows. In section \ref{sec:proof}, we present the main propositions concerning the short time asymptotics of linear dynamical systems whose time evolution is governed by a possibly time dependent but sparse matrix. In section \ref{sec:discussion}, we apply these statements to closed and open CTQWs and illustrate our results by various case studies including chiral walks. Conclusion and future direction of research are given in section \ref{sec:conclusion}.

\section{Main mathematical results} \label{sec:proof}

\subsection{The main theorem}

Throughout this  section $\mathcal{H}$ will denote a finite-dimensional Hilbert space with an orthonormal basis $\{ \ket{v} \}_{v \in \mathcal{V}}$  labeled by the vertices of a graph $\mathcal{G}=(\mathcal{V},\mathcal{E})$ with edge set $\mathcal{E}$. The graph is assumed to be simple and  directed. For each distinct vertices $n,m$ of the graph $\mathcal{G}$, we denote the set of the shortest, directed paths connecting $n$ to $m$ by $\mathcal{P}(n,m)$. If $p$ is a path in $\mathcal{P}(n,m)$ of length $d$, then it can be represented by a sequence of vertices $p_0,\dots,p_d$ with $p_0=n$, $p_d=m$ and the edges $(p_k,p_{k+1})\equiv p_k\rightarrow p_{k+1}$ formed by the consecutive members of $p_0,\dots,p_d$ are just the edges of $p$. 

We consider a continuous family of linear operators $[0,T] \ni t \mapsto M(t) \in \mathcal{B}(\mathcal{H})$ satisfying the property $\bra{m} M(t)\ket{n}\not\equiv 0$ on $[0,T]$ if and only if the directed edge $(n,m)$ is a member of $\mathcal{E}$. In that case, we say that $\mathcal{G}$ is the graph of $M(t)$. Given distinct vertices $n$ and $m$, and a shortest path $p\in\mathcal{P}(n,m)$ of length $d$, we define the corresponding path amplitude $\Phi_p[M(t)]$ as
\begin{eqnarray}
&&\Phi_p[M(t)]=\int_{0}^{t}\mathrm{d}s_d\cdots\int_{0}^{s_2}\mathrm{d}s_1\, \bra{p_d}M(s_d)\ket{p_{d-1}}\times\cdot\cdot\nonumber\\
&&\qquad\qquad\qquad\qquad\times\cdot\cdot \bra{p_1}M(s_1)\ket{p_0}.
\label{eq:phidef}
\end{eqnarray}
If $A$ is a matrix operating on $\mathcal{H}$, its norm is defined through 
\begin{equation}
\|A\|=\max_{\psi\neq 0}\frac{\|A\psi\|}{\|\psi\|},
\label{eq:normdef}
\end{equation}
where $\|\psi\|=\sqrt{\braket{\psi}{\psi}}$. Note that the norm satisfies the inequality 
\begin{equation}
\|A+cB\|\leq \|A\|+|c|\|B\|
\label{eq:norm1}
\end{equation}
for any complex $c$, and has the submultiplicative property
\begin{equation}
\|AB\|\leq \|A\|\,\|B\|.
\label{eq:norm2}
\end{equation}
Let $\tau_T$ denote the reciprocal of the maximum among the norms of $\|M(t)\|$ if $t$ runs from zero to $T$:
\begin{equation} 
\tau^{-1}_T =\max_{0 \le t \le T} \| M(t)\|.
\label{eq:tau}
\end{equation}

We can now state and prove the main theorem on short time asymptotics.\par
\vspace{3mm}
\noindent \textbf{Proposition 1.} \textit{The solution of the matrix differential equation
\begin{equation}
\frac{d}{dt}X(t) = M(t) X(t) \, , \; \; X(0)=\unity \label{eq:diff_a1}
\end{equation} 
satisfies the inequality
\begin{equation}
\Bigl| \bra{m} X(t) \ket{n} - \sum_{p \in \mathcal{P}(n,m)} \Phi_p[M(t)] \, \Bigr| \le \mathrm{e}^{t/\tau_T}  \frac{ (t/\tau_T)^{d(n,m){+}1}}{(d(n,m){+}1)!} \label{eq:main_a1}
\end{equation}
for all $n,m \in V$ of distance $d(n,m)$. Here, the sum goes over the set of shortest paths $\mathcal{P}(n,m)$ running from $n$ to $m$ in $\mathcal{G}$, and 
$\Phi_p[M(t)]$ is defined in Eq.~\eqref{eq:phidef}.}\par
\vspace{2mm}
Before proving the statement, some remarks should be added. First, note that $\mathrm{i}M(t)$ in the proposition is not necessarily hermitian. Indeed, it can be any square matrix. This fact gives the opportunity to apply the statement also to Lindbladian dynamics in section \ref{sec:discussion}. The characteristic measure of the short time dynamics is $\tau_T$, that the approximation contained in Eq.~\eqref{eq:main_a1} is informative only whenever $t$ is less than $\tau_T$. For time-independent generators, $\tau_T$ is independent of $T$. Moreover, $\tau_T$ does not depend on a complex prefactor of modulus one multiplying $M(t)$. Since every CTRW taking place on a symmetric weighted graph has a corresponding CTQW with the same generator but multiplied by $-\mathrm{i}$, the scales of the short time asymptotics are necessarily identical. In the case of chiral CTQW, the appearance of the path amplitudes $\Phi_p[M(t)]$ in Eq.~\eqref{eq:main_a1} results in interference patterns with which CTQW obtains a richer structure as compared to CTRW, where the amplitudes are always positive.  

\vspace{2mm}
\noindent \textbf{Proof.} As $t \mapsto M(t)$ is a continuous map, the solution of the differential equation \eqref{eq:diff_a1} can be written as the sum of the Dyson series
\begin{equation}
X(t)= \sum_{N=0}^{\infty} \int_{0}^{t}  {\rm d} s_{N} \cdots \int_{0}^{s_2} {\rm d} s_{1}  \, M(s_N) \cdots M(s_1). \label{eq:pic_1}
\end{equation}
Let $d$ be the graph distance between nodes $n$ and $m$. Then, for any $k < d$ and for any $0\leq s_1, s_2, \ldots s_k\leq T$ the identity $\bra{m} M(s_k) \cdots M(s_1)  \ket{n}=0$ holds. Thus, when calculating the entry $[X(t)]_{mn}$, the Dyson series reduces to  
\begin{eqnarray}
&&\bra{m} X(t)  \ket{n} \nonumber \\
&&\ =\sum_{N=0}^{\infty} \int_{0}^{t}  {\rm d} s_{N} \cdots \int_{0}^{s_2} {\rm d} s_{1} \bra{m}  M(s_N) \cdots M(s_1) \ket{n}  \nonumber \\
&&\ =\int_{0}^{t}  {\rm d} s_{d} \cdots \int_{0}^{s_2} {\rm d} s_{1} \bra{m}  M(s_d) \cdots M(s_1) \ket{n} \nonumber \\
&&\qquad +\sum_{N=d+1}^{\infty} \int_{0}^{t}  {\rm d} s_{N} \cdots \int_{0}^{s_2} {\rm d} s_{1} \bra{m}  M(s_N)\cdots\nonumber\\
&&\qquad \qquad  \cdots M(s_1) \ket{n}. \label{eq:pic_2}
\end{eqnarray}
For any linear operator $A$, one has $\| A \| \ge  |\bra{m} A \ket{n}|$, so we can bound each term in Eq.~\eqref{eq:pic_2} as
\begin{eqnarray}
&&\left|\int_{0}^{t}  {\rm d} s_{N} \cdots \int_{0}^{s_2} {\rm d} s_{1} \bra{m}  M(s_N) \cdots M(s_1) \ket{n}  \right|   \nonumber \\
&&\ \ \leq\int_{0}^{t}  {\rm d} s_{N} \cdots \int_{0}^{s_2} {\rm d} s_{1}  \left| \bra{m}  M(s_N) \cdots M(s_1) \ket{n}  \right|   \nonumber \\
&&\ \ \leq \int_{0}^{t}  {\rm d} s_{N} \cdots \int_{0}^{s_2} {\rm d} s_{1}  \|  M(s_N) \cdots M(s_1) \|  \nonumber \\
&&\ \ \leq \int_{0}^{t}  {\rm d} s_{N} \cdots \int_{0}^{s_2} {\rm d} s_{1}  \|  M(s_N)\| \cdots \|M(s_1) \|  \nonumber \\
&&\ \ \leq \int_{0}^{t}  {\rm d} s_{N} \cdots \int_{0}^{s_2} {\rm d} s_{1}\,  \frac{1}{\tau^N_T} =\frac{(t/\tau_T)^N}{N!}. 
\end{eqnarray}
Therefore, 
\begin{eqnarray}
&&\left|\sum_{N=d+1}^{\infty}\int_{0}^{t}  {\rm d} s_{N} \cdots \int_{0}^{s_2} {\rm d} s_{1} \bra{m}  M(s_N) \cdots M(s_1) \ket{n}  \right|   \nonumber \\
&&\ \ \leq\sum_{N=d+1}^{\infty}\frac{(t/\tau_T)^N}{N!}\leq \frac{(t/\tau_T)^{d+1}}{(d+1)!}\mathrm{e}^{\xi}\nonumber\\
&&\ \ \leq \frac{(t/\tau_T)^{d+1}}{(d+1)!}\mathrm{e}^{t/\tau_T}, \label{eq:remainder_1}
\end{eqnarray}
where we used Taylor's theorem with the Lagrange form of the remainder, which holds with a suitably chosen $\xi \in [0, t/\tau_T]$. This implies 
\begin{eqnarray}
&& \Biggl| \bra{m} X(t)  \ket{n} \nonumber\\
&&\quad{-}   \int_{0}^{t}  {\rm d} s_{d} ... \int_{0}^{s_2} {\rm d} s_{1} \bra{m}  M(s_d) \cdots M(s_1) \ket{n} \Biggr|\nonumber\\ 
&&\quad\quad\quad \le e^{t /\tau_T} \frac{(t/\tau_T)^{d+1}}{(d+1)!}.
\label{eq:aux}
\end{eqnarray}
Now, let us perform the expansion 
\begin{eqnarray}
&& [M(s_d) \cdots M(s_1)]_{mn}=\nonumber\\
&&\qquad=\sum_{k_1, \dots,k_{d-1}} [M(s_d)]_{m k_{d-1}} \cdots  [M(s_1)]_{k_1 n}. 
\end{eqnarray}
It is clear that only those indices contribute in the above sum for which $(n, k_{1}, k_{2} \ldots, k_{d-1},  m)$ forms a path in $\mathcal{G}$ connecting $n$ to  $m$. This means that one can replace the above sum over vertex sets to a sum over the path-set $\mathcal{P}(n,m)$:
\begin{eqnarray}
&& [M(s_d) \cdots M(s_1)]_{mn}=\nonumber\\
&&\qquad =\sum_{p \in \mathcal{P}(n,m)} [M(s_d)]_{m p_{d-1}} \cdots  [M(s_1)]_{p_1 n}. \label{eq:amplitude_1}
\end{eqnarray}
Inserting this into Eq.~\eqref{eq:aux}, we arrive to Eq.~\eqref{eq:main_a1}. $\blacksquare$\par
\vspace{3mm}

\subsection{Improvement of the timescale}

The main drawback of Proposition 1 is the appearance of the norms $\|M(t)\|$. Choosing the Hilbert space basis $\ket{n}$, and assuming that $M$ is constant in time, then splitting $M$ to a sum of diagonal and off-diagonal parts (which is the case for instance in tight-binding models) and varying only the diagonal entries affects the timescale $\tau$ dramatically. However, using time-dependent perturbation theory, more can be said than what Eq.~\eqref{eq:main_a1} would allow. Let $M=V+\hat{M}$ be an arbitrary square matrix with diagonal part $V$ and off-diagonal part $\hat{M}$. Let $\lambda\geq 0$ be the smallest real satisfying $\Re(V-\lambda)\leq 0$. Let $A$ be the adjacency matrix obtained by setting all non-zero entries of $\hat{M}$ to one. The graph $\mathcal{G}=(\mathcal{V},\mathcal{E})$ described by $A$ is simple but directed: the edge $(n,m)$ with tail $n$ and head $m$ is a member of $\mathcal{E}$ if and only if $\bra{m}A\ket{n}=1$. Define $\hat{M}(t)$ as 
\begin{equation}
\hat{M}(t)=\exp(-Vt)\hat{M}\exp(Vt).
\end{equation}
\vspace{3mm}
\noindent\textbf{Proposition 2.} \textit{The following inequality holds:
\begin{eqnarray}
&&\Bigl| \bra{m} \exp(Mt) \ket{n} - \mathrm{e}^{V_mt}\sum_{p \in \mathcal{P}(n,m)} \Phi_p[\hat{M}(t)] \, \Bigr|\nonumber\\
&&\qquad \le \mathrm{e}^{t/\tau} \mathrm{e}^{\lambda t}  \frac{ (t/\tau)^{d(n,m){+}1}}{(d(n,m){+}1)!} \label{eq:main_a2},
\end{eqnarray}
for all $n,m \in \mathcal{V}$, where 
\begin{equation}
\tau^{-1}=\|A\|\,\|\hat{M}\|_{\max}=\|A\|\,\max_{n,m}|\bra{m}\hat{M}\ket{n}|,
\label{eq:tau_def2}
\end{equation}
$V_m=\bra{m}V\ket{m}$ and $\mathcal{P}(n,m)$ is the set of shortest directed paths connecting $n$ to $m$ in $\mathcal{G}$ of length $d(n,m)$.}\par
\vspace{2mm}
\noindent\textbf{Proof.} Define $\hat{V}=V-\lambda$. Note that 
\begin{equation}
\exp(Mt)=\exp(\lambda t)\exp(\hat{V}t)X(t),
\end{equation}
where $X(t)$ is the solution to 
\begin{equation}
\frac{d}{dt}X(t)=\hat{M}(t)X(t),\qquad X(0)=\mathbbm{1}\label{eq:main_a2}.
\end{equation}
Also note 
\begin{equation}
\hat{M}(t)=\exp(-Vt)\hat{M}\exp(Vt)=\exp(-\hat{V}t)\hat{M}\exp(\hat{V}t).
\end{equation}
Let $s_{N+1}=t$ and $s_{0}=0$. Then, the $N$th order term of the Dyson series of Eq.~\eqref{eq:main_a2} multiplied by $\exp(\hat{V}t)$ is of the form 
\begin{equation}
Y_{N}(t)=\int_{0}^t\mathrm{d}s_N\cdots \int_{0}^{s_2}\mathrm{d}s_1\left[\prod_{k=1}^{N}\mathrm{e}^{\hat{V}(s_{k+1}-s_{k})}\hat{M}\right]\mathrm{e}^{\hat{V}(s_1-s_0)}.
\end{equation}
Since $\Re(V-\lambda)\leq 0$ and $0=s_0\leq s_1\leq \cdots\leq s_N\leq s_{N+1}=t$ holds, we have the following upper bounds:
\begin{eqnarray}
|\bra{u}\mathrm{e}^{\hat{V}(s_{k+1}-s_k)}\hat{M}\ket{v}|&\leq& |\bra{u}\hat{M}\ket{v}|\leq \|\hat{M}\|_{\mathrm{max}}A_{uv}\nonumber,\\
|\bra{u}\mathrm{e}^{\hat{V}(s_{1}-s_0)}\ket{v}|&\leq&\delta_{uv},
\end{eqnarray}
which hold for any two vertices $u$ and $v$ of $\mathcal{G}$, so we can write
\begin{eqnarray}
&&|\bra{m}Y_N(t)\ket{n}|\nonumber\\
&&\quad\leq \|\hat{M}\|_{\mathrm{max}}^N\int_{0}^t\mathrm{d}s_N\cdots \int_{0}^{s_2}\mathrm{d}s_1 \sum_{k_1}\cdots\sum_{k_N}A_{m,k_N}\cdots A_{k_1,n}\nonumber\\
&&\quad=\|\hat{M}\|_{\mathrm{max}}^N\bra{m}A^N\ket{n}\frac{t^N}{N!}.
\label{eq:Hest}
\end{eqnarray}
Therefore, since $|\bra{m}A^N\ket{n}|\leq \|A\|^N$ holds, we find
\begin{eqnarray}
|\bra{m}\mathrm{e}^{\lambda t}Y_N(t)\ket{n}|&\leq& \mathrm{e}^{\lambda t}\frac{\bigl(\|\hat{M}\|_{\mathrm{max}}\|A\|t\bigr)^N}{N!}\\\nonumber
&=& \mathrm{e}^{\lambda t}\frac{1}{N!}\left(\frac{t}{\tau}\right)^N,
\end{eqnarray}
where $\tau$ is given in Eq.~\eqref{eq:tau_def2}. From this point, the arguments of the proof of Proposition 1 can be repeated to obtain
\begin{eqnarray}
&&\left|\bra{m}\exp(Mt)\ket{n}-\mathrm{e}^{\lambda t}\bra{m}Y_{d(n,m)}\ket{n}\right|\nonumber\\
&&\quad=\Bigl|\bra{m}\exp(Mt)\ket{n}-\mathrm{e}^{V_mt}\bra{m}\sum_{p \in \mathcal{P}(n,m)}\Phi_p[\hat{M}(t)]\ket{n}\Bigr|\nonumber\\
&&\quad\leq \frac{\mathrm{e}^{\lambda t}}{(d(n,m)+1)!}\left(\frac{t}{\tau}\right)^{d(n,m)+1},
\end{eqnarray}
which proves the statement. $\blacksquare$\par
\vspace{2mm}
It is a well-known fact that the largest eigenvalue of the adjacency matrix $A$ of a simple, undirected graph $\mathcal{G}$ is also largest in magnitude, and is bounded by the highest degree $\mathrm{d}_\mathrm{max}(\mathcal{G})$ of $\mathcal{G}$ from above. That is, when $\hat{M}$ admits the property $\bra{u}\hat{M}\ket{v}=0$ if and only if $\bra{v}\hat{M}\ket{u}=0$, then $A$ is symmetric, so
\begin{eqnarray}
\|A\|=\lambda_{\mathrm{max}}(A)\leq \mathrm{d}_\mathrm{max}(\mathcal{G}).
\label{eq:Aest}
\end{eqnarray}
A particular example is the tight-binding model, taking place on the simple, undirected graph $\mathcal{G}$ with adjacency matrix $A$. Then, $M$ can be replaced in Proposition 2 by $-\mathrm{i}H=-\mathrm{i}(V+A)$ to obtain:
\begin{eqnarray}
&&\Bigl| \bra{m} \exp(-\mathrm{i}Ht) \ket{n} - \mathrm{e}^{-\mathrm{i}V_mt}\sum_{p \in \mathcal{P}(n,m)} \Phi_p[\hat{H}(t)] \, \Bigr|\nonumber\\
&&\qquad \le \mathrm{e}^{t/\tau}  \frac{ (t/\tau)^{d(n,m){+}1}}{(d(n,m){+}1)!},
\end{eqnarray}
where
\begin{equation} 
\tau^{-1}=\mathrm{d}_\mathrm{max}(\mathcal{G})\,\max_{n\neq m}|\bra{n}\hat{H}\ket{m}|.
\end{equation}
\par

\section{Application of the results}\label{sec:discussion}
\subsection{Comparison of CTRW and CTQW} \label{ssec:CTRW}
In order to compare the short time asymptotics of the probabilistic and unitary versions of continuous 
time walks, we fix a simple, undirected graph $\mathcal{G}=(\mathcal{V},\mathcal{E})$, containing no self-loops. Using the adjacency matrix 
$A$ and the degree matrix $D$, the CTRW dynamics is generated by the graph Laplacian \cite{chung1997spectral, telcs2006art} $L=D-A$,  that is, if 
$u,v\in \mathcal{V}$ are arbitrary vertices, then the conditional probability of observing the walker at vertex $u$ 
if its initial position was $v$ is
\begin{equation}
p_{\mathrm{R}}(u,t|v)=\bra{u}\exp(-Lt)\ket{v}.
\label{eq:ctrw}
\end{equation} 
The unitary walk on the same graph is generated by $-\mathrm{i}L$ with transition probabilities given by
\begin{equation}
p_{\mathrm{Q}}(u,t|v)=|\bra{u}\exp(-\mathrm{i}Lt)\ket{v}|^2. 
\end{equation}
Since $\|L\|=\|-\mathrm{i}L\|$, the norm of the generators which define the timescale $\tau=\|L\|^{-1}$ of the short-time regime are equal, the two dynamics defined above and the hitting probabilities are naturally comparable. 
We choose the graph $\mathcal{G}$ to be a binary tree depicted in Fig.~\ref{fig:fig1}. It is clearly visible that the numerical results fit very well to the theoretical curves in the time horizon $t<\tau$ both in case of CTQW and CTRW. The only exception is the $0 \rightarrow 1$ transition, where the error of the approximative formula becomes significant already for $t > \tau/2$. However, this is easily understandable  if one notes that in that case the denominator of the error bound appearing in Eq.~\eqref{EQ3} becomes comparable to the numerator.
\newpage
\onecolumngrid
\begin{center}
\centering
\begin{figure}
\includegraphics[width=1\textwidth]{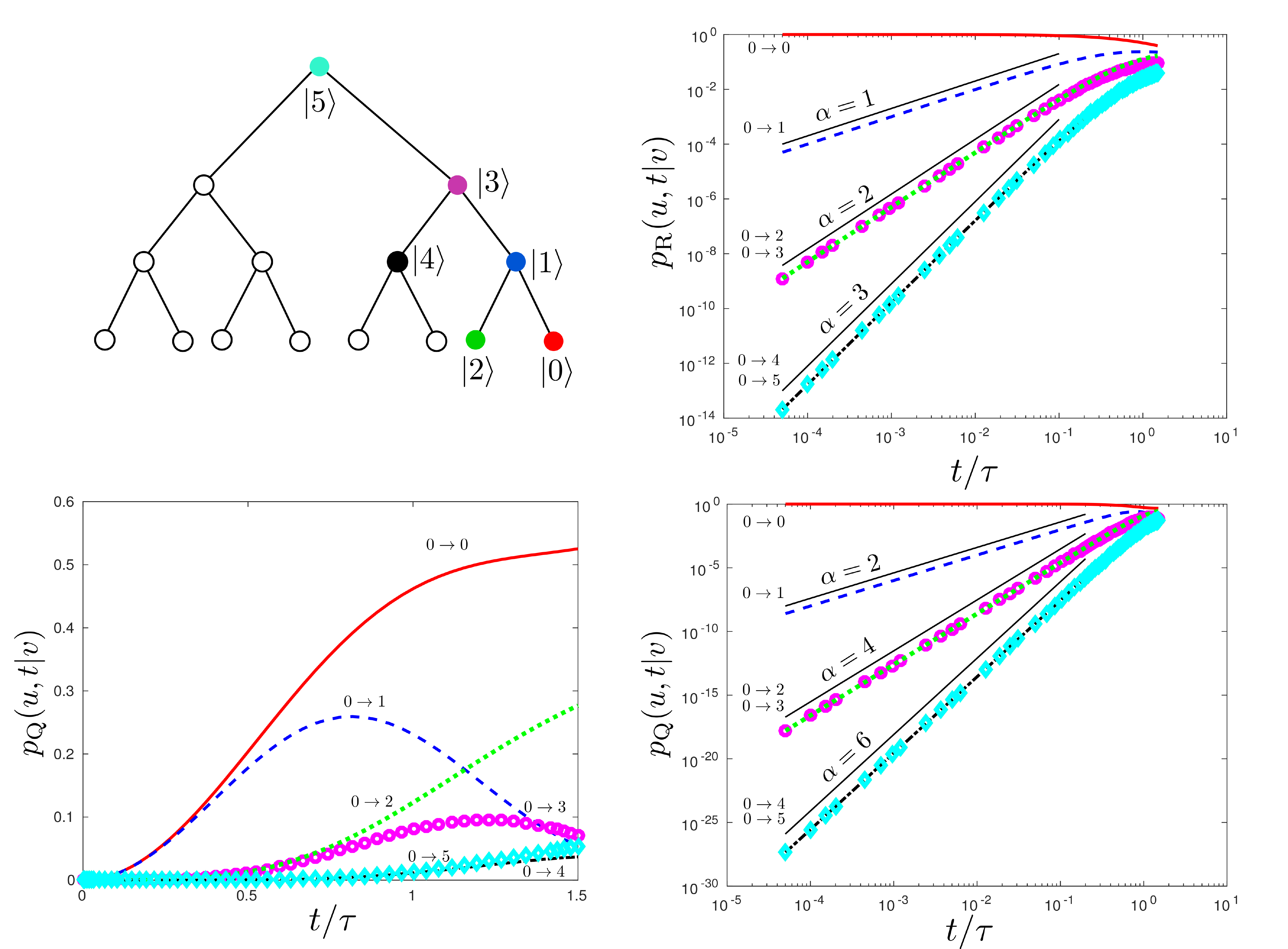}
\caption{Comparison of the CTRW and the CTQW taking place on the graph depicted in the top left corner of the figure. Transition and hitting probabilities--as 
functions of time--have been calculated between vertex $\ket{0}$ on the one hand and vertices $\ket{0}$, $\ket{1}$, $\ket{2}$, $\ket{3}$, $\ket{4}$, $\ket{5}$ on the other.
 Beside the results of the numerical calculations, log-log plots depict the predictions of Eq.~\eqref{EQ1} and Eq.~\eqref{EQ2} with the corresponding exponents $\alpha=d$ in the case of CTRW, and $\alpha=2d$ in the case of CTQW respectively, $d$ being the graph distance of the nodes. Note that these theoretical curves--for the sake of better comparison--have been slightly shifted in the vertical direction. Whenever two time series overlap on the log-log plot, dashed lines represent $p_\mathrm{X}(2,t|0)$ and $p_\mathrm{X}(4,t|0)$, while circles and diamonds represent $p_\mathrm{X}(3,t|0)$ and $p_\mathrm{X}(2,t|0)$ respectively, either if $\mathrm{X}$ denotes $\mathrm{R}$ or $\mathrm{Q}$. For a sake of better comparison, the linearly scaled diagram in the bottom left corner of the figure contains the numerics of $1-p_{\mathrm{Q}}(0,t|0)$ instead of $p_{\mathrm{Q}}(0,t|0)$. (Color online.)}
\label{fig:fig1}
\end{figure}
\end{center}
\twocolumngrid

\subsection{CTQWs  with arbitrary on-site potential } \label{ssec:TightBinding}
In order to demonstrate the universality of the short time asymptotics  in tight-binding models, we consider Hamiltonians of the form $H=A+V$, where $V$ is a diagonal matrix, called the on-site potential. The hitting probabilities are
\begin{equation}
p_{\mathrm{TB}}(u,t|v) = |\bra{u}\exp(-\mathrm{i}(A+V)t)\ket{v}|^2.
\end{equation}
Consider the graph that has been introduced in the previous subsection \ref{ssec:CTRW}. We choose the on-site potentials from an ensemble of independent, identically distributed Gaussian random variables with mean zero and unit variance. Fig.~\ref{fig:fig2} illustrates the transition probabilities between vertices of different distances.  
The time series depicted in Fig.~\ref{fig:fig2} has been obtained by first calculating the full time 
series of the hitting probabilities between fixed sites $v$ and $u$ for $75$ different random realizations 
of $V$. If the index $\alpha=1\dots 75$ marks the different realizations of the on-site potential, 
then these numerical calculations resulted in sequences of pairs $(t_{k}/\tau_\alpha,p^{(\alpha)}_{uv}(t_k/\tau_\alpha))$, $t_k/\tau_\alpha$, 
$k=1\dots 75$ varying between $0.5\times 10^{-5}$ and $1.5$. Here, $\tau_\alpha=\|A+V_\alpha\|$. After that, the diagonal sequence 
$(t_{\alpha}/\tau_\alpha,p^{(\alpha)}_{uv}(t_\alpha/\tau_\alpha))$ has been plotted. The figure provides strong  evidence of the 
independence of the short time asymptotics from the on-site potential. 

\begin{center}
    \centering
    \begin{figure}[t]. 
    \includegraphics[width = 0.52\textwidth]{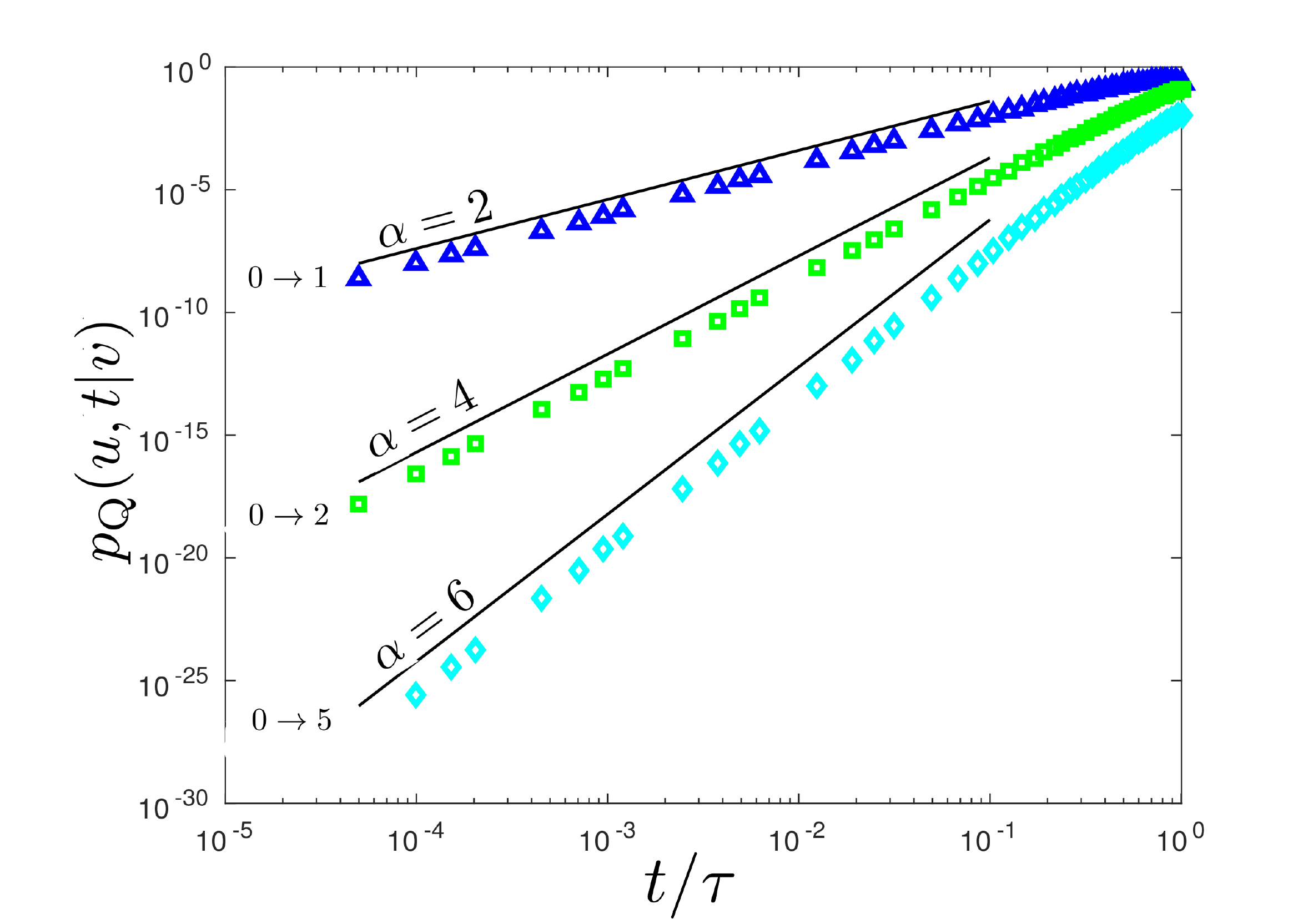}
    \caption{Universality in tight-binding models with Gaussian distributed on-site potentials. Detailed explanation in the main text. (Color online.)}
    \label{fig:fig2}
    \end{figure}
\end{center}

\subsection{Chiral Quantum Walks} \label{ssec:chiral}

Next, we discuss the short time properties of chiral walks \cite{zimboras2013quantum,lu2016chiral}.  These walks are defined by modifying the

\begin{center}
    \centering
    \begin{figure}[!ht]
    \includegraphics[width = 0.44\textwidth]{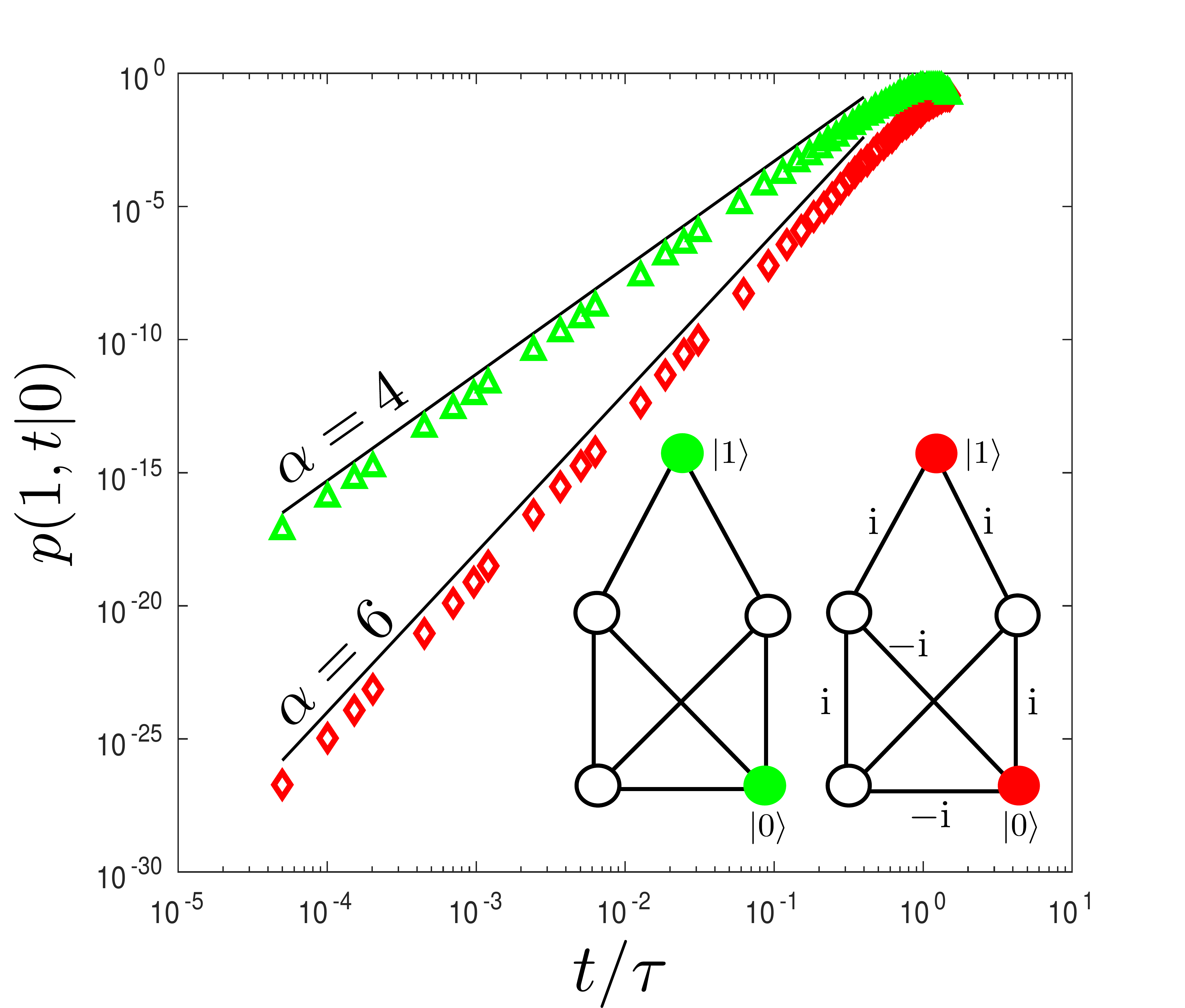}
    \caption{Comparison of time-reversal symmetric and chiral quantum walks on the graph depicted in the figure. Detailed explanation in the main text. (Color online.)}
    \label{fig:fig3}
    \end{figure}
\end{center}

adjacency matrix of a graph $ \mathcal{G}$ by assigning a complex phase to a transition $\ket{n} \to \ket{m}$ allowed by the 
adjacency matrix, and the conjugate phase to the transition $\ket{m} \to \ket{n}$, i.e.,  by defining the Hamiltonian 
\begin{equation}\label{eqn:cqw} 
  H_{ch} = \sum_{\{n,m \} \in \mathcal{E}} e^{i \theta_{nm}} \ket{n} \! \bra{m} +    e^{-i \theta_{nm}}\ket{m} \! \bra{n} .
\end{equation}

Chiral quantum walks offer a flexible way to engineer transport properties of quantum networks. For example, while for a non-chiral CTQW the transition probabilities satisfy the time-reversal and reflection symmetries, i.e., $p(x,t|y)= p(x, -t |y)$ and $p(x,t|y)= p(y, t |x)$,  for chiral walks these may be broken and only the composition of these symmetries are satisfied, $p(y,t|x)= p(x,-t|y)$. This freedom has been used to direct, enhance or suppress transport by tuning the complex phases \cite{zimboras2013quantum, lu2016chiral, manzano2018harnessing, wong2015quantum}.

Similarly, it turns out that chiral walks display also highly adjustable short-time properties compared with their non-chiral counter parts. By varying the strength of the diagonal potential terms or the off-diagonal hopping terms of non-chiral CTQWs, one cannot change the leading exponent of $t$ in the short-time expansion of the transition probabilities as discussed in the previous subsection. Contrary to this, one can (in case of some network topologies) change the leading exponent by adjusting  the phase factors in a chiral walk Hamiltonian. This can be easily shown: Consider a chiral quantum walk Hamiltonian on $\mathcal{G}$ which we divide into a diagonal and an off-diagonal term, $H=D+O$,  where exactly those entries $O_{kl}$ of the off-diagonal term are non-zero for which the nodes $k$ and $l$ are connected. As discussed in Section \ref{sec:proof},  one can show for the transition probability that 
\begin{eqnarray}
&p(m, t | n )= \frac{|\ell(m,n)|^2}{(d(n,m)!)^2}t^{2d(n,m)}+\mathcal{O}(t^{2d(n,m)+1}),  \\
& \ell(n,m) = \sum_{p \in \mathcal{P}(n,m)} \Phi_p[O],
\end{eqnarray}
where the sum goes over the different shortest paths $\mathcal{P}(n,m)$ from $n$ to $m$.  If we tune the phases of the off-diagonal entries $O_{kl}$ to be positive reals, then $ \ell(n,m) $ is non-zero, and the leading exponent is $2d(n,m)$. However, for certain geometries, we can choose the phases of these entries in such a way that the sum over different paths cancel each other and the first non-vanishing leading term will then have a leading exponent larger than $2d(n,m)$. The effect of such a cancellation on a specific graph $\mathcal{G}$ is illustrated on Fig.~\ref{fig:fig3}. Note that the particular graph we choose in this case could not be a tree graph, since the phases then can be transformed out yielding a  non-chiral CTQW with the same transition probabilities as the original chiral walk \cite{zimboras2013quantum} (see also Appendix ~\ref{app:A}). It is clear that with the specified arrangement of complex phases with respect to the transition  $\ket{0} \to \ket{1}$ the leading exponent  is $6$, contrary to the non-chiral case when it is $4$.

\subsection{Time-dependent Hamiltonian dynamics}

To study CTQW in time-dependent tight-binding models, let us consider a time-dependent Hamiltonian of the form 
\begin{equation}
    H(t) = \Lambda^{+}(t)\,A\,\Lambda(t),
\label{eq:eq_time_dep}
\end{equation}
where $A$ is the adjacency matrix of a simple, undirected graph, containing no self-loops and $\Lambda(t)$ is a family of unitary matrices, not commuting with $A$ for all time instances. Note that, for any choice of
$\Lambda(t)$ unitarity guarantees that $\tau_T$ introduced in Eq.~\eqref{eq:tau} is 
determined solely by $A$:
\begin{eqnarray}
    &&\|\Lambda^{+}(t)A\Lambda(t)\|= \max_{\psi\neq 0}\frac{\|\Lambda^{+}(t)A\Lambda(t)\psi\|}{\|\psi\|}\nonumber\\
    &&\quad =\max_{\psi\neq 0}\frac{\|A\Lambda(t)\psi\|}{\|\Lambda(t)\psi\|}=\|A\|.
\end{eqnarray}

\begin{center}
    \centering
    \begin{figure}[t]
    \includegraphics[width = 0.46\textwidth]{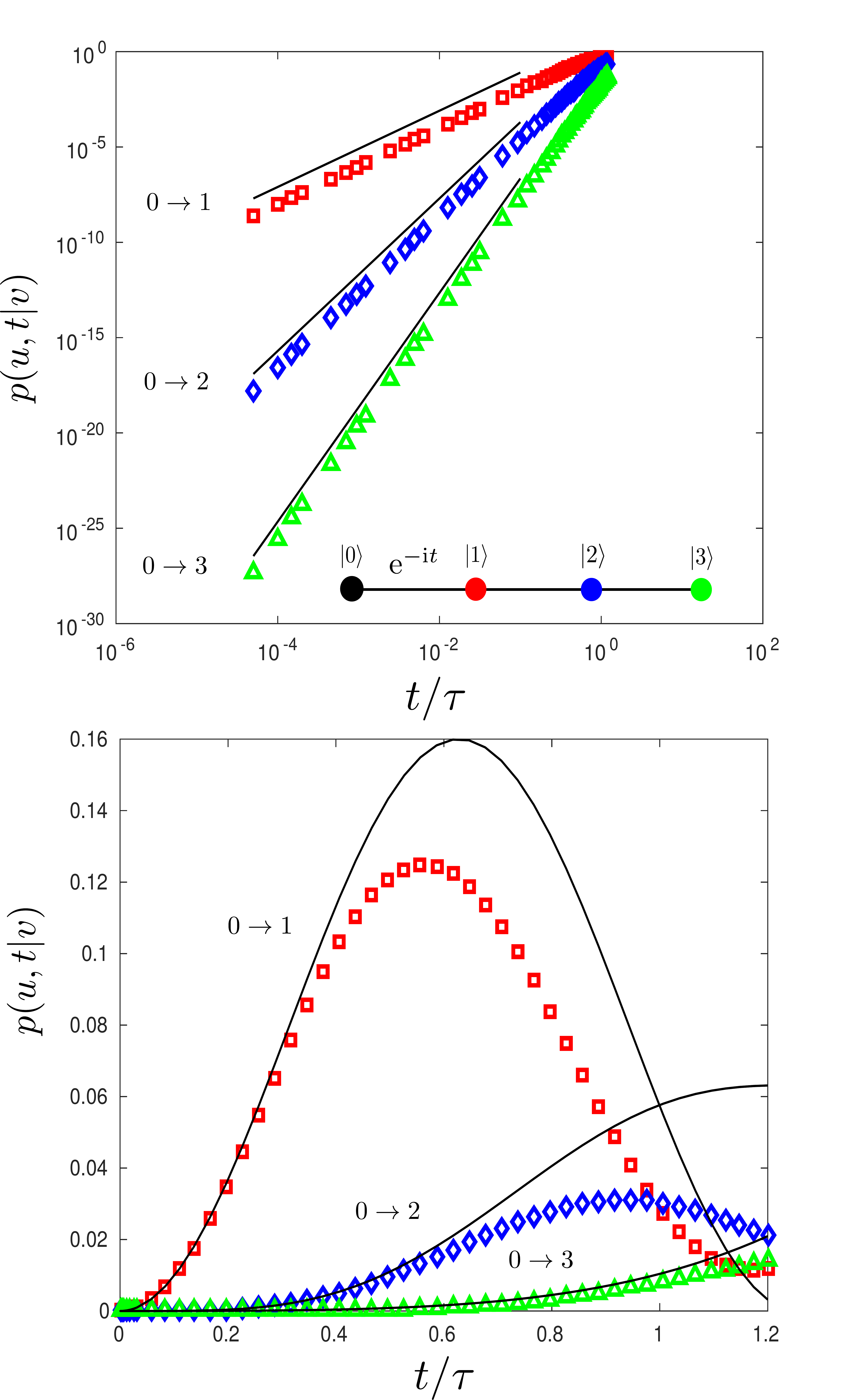}
    \caption{Comparison of numerical and theoretical results on time-dependent tight-binding model depicted in the figure. Black lines represent, while the marks represent the numerical calculation for the transition probabilities with different path length, when $\Omega = 1$. Theoretical curves are evaluated by the equation \ref{eq:diff_a1}. (Color online.)}    
    \label{fig:fig4}
    \end{figure}
    \end{center}

We choose the particular case when $\Lambda(t)=\exp(-\mathrm{i}\Omega t)$, $\Omega$ being a real diagonal matrix. Due to $[H(t),H(t')]\neq 0$ for distinct time instances $t$ and $t'$, the resulting time evolution of Eq.~\eqref{eq:eq_time_dep} can be significantly complicated. This is always the case, even if $A$ represents a tree, which cannot support a non-trivial chiral walk. To illustrate the effect of the appearance of $\Lambda(t)$ in Eq.~\eqref{eq:eq_time_dep}, $A$ is chosen such that it corresponds to a linear chain of length $n$ whose nodes are labeled in linear order from $0$ to $n$, while $\Lambda(t)=\mathrm{diag}(\mathrm{e}^{-\mathrm{i}\Omega t},1,\dots,1)$. A short calculation shows that the short time asymptotics of the transition amplitudes $\bra{v} U(t)\ket{0}$ are
\begin{equation}
\bra{v} U(t)\ket{0} = \dfrac{1}{\Omega^d}\bigg(-\sum_{u=0}^{v-1} (-\mathrm{i}\Omega)^u \dfrac{t^u}{u!} + \mathrm{e}^{-\mathrm{i}\Omega t} \bigg) + \mathcal{O}(t^{d+1}).
\label{eq:eq_approx_chain}
\end{equation}
Fig.~\ref{fig:fig4} shows the comparison of the exact numerical calculations and the approximative formula of Eq.~\eqref{eq:eq_approx_chain}, which corresponds to a chain of three links, one of them admitting a rotating phase. Fig.~\ref{fig:fig4} shows that the theoretical curves fit well in the time horizon $t<0.5\tau$. In the time horizon $0.5\tau < t$, the error of the approximative formula becomes significant as we have seen in the previous section.

According to Proposition 2, when  $\Omega$ is replaced by $V$, the potential matrix of a tight-binding model, one can think about $H(t)$ in Eq.~\eqref{eq:eq_time_dep} as the Hamiltonian of the system in the interaction picture. This allows one to extend the validity of the short time approximation from the time horizon defined by  $\tau_V=\|A+V\|^{-1}$ to that of $\tau=\|A\|^{-1}$, which is usually greater than $\tau_V$ according to Gershgorin's circle theorem. One consequence of this is the claim that localization needs more time than $\tau_V$ to develop: at the timescale  $\tau_V$, Eq.~\eqref{EQ3} implies the constant increase of the transition probabilities in time. Here, we show that such a behavior persists also at the timescale $\tau$.

Let  $\mathcal{G}$ be a simple, undirected graph and assume that the non-vanishing entries of $V$ are i.i.d Gaussian random variables centered around the origin and have unit variance. Let $n,m$ be two non-identical nodes of $\mathcal{G}$. We show that the disorder average of the approximative transition probability from $n$ to $m$ increases monotonically. Assume that the distance of nodes $n$ and $m$ is $d$ and let  $n=p_0,p_1,\dots,p_{d-1},p_d$ define a shortest path connecting $n$ to $m$ within $\mathcal{G}$. Define $s(t)$ as a tuple  of positive reals containing $d+1$ elements with $0=s_{0}\leq s_0\leq \cdots \leq s_d=t$ and let $\Sigma_t$ be the set of such tuples. For any vertex $w$, tuple $s(t)$ and path $p$ of length $d$ let 
\begin{equation}
R(p,w,s(t))=\sum_{k=0}^{d}\delta(p_k,w)(s_{k+1}-s_{k}),
\end{equation}
where $\delta$ stands for Kronecker's delta function. Then, using Proposition 2 we obtain
\begin{eqnarray}
&&\big\langle p_{\mathrm{TB}}(m,t|n)\big\rangle_{V}=\sum_{p\in\mathcal{P}(n,m)}\sum_{p'\in\mathcal{P}(m,n)}\int_{\Sigma_t}\mathrm{d}s\int_{\Sigma_t}\mathrm{d}s'\nonumber\\
&&\ \times \Big\langle \prod_{w\in V}\exp\big{\{}-\mathrm{i}V_{w}\big[R(p,w,s(t))-R(p',w,s'(t))\big]\big{\}}\Big\rangle_V\nonumber\\
&&\qquad+\mathcal{O}(t^{2d(n,m)+1}).
\end{eqnarray} 
Define $\Phi(x)$ as the generating function of the Gaussian distribution centered around the origin and having unit variance:
\begin{eqnarray}
\Phi(x)=\int_{-\infty}^{\infty}\frac{1}{\sqrt{2\pi}}\mathrm{e}^{-V^2/2}\mathrm{e}^{-\mathrm{i}xV}\,\mathrm{d}V=\mathrm{e}^{-2x^2},
\end{eqnarray}
and note that 
\begin{eqnarray}
&&\Big\langle \prod_{w\in V}\exp\big[-\mathrm{i}V_{w}\big{\{}R\big(p,w,s(t)\big)-R\big(p',w,s'(t)\big)\big{\}}\big] \Big\rangle_V\nonumber\\
\quad&&=\prod_{w\in V}\Big\langle \exp \big[-\mathrm{i}V_{w}\big{\{}R\big(p,w,s(t)\big)-R\big(p',w,s'(t)\big)\big{\}}\big]\Big\rangle_V\nonumber\\
\quad&&=\prod_{w\in V} \Phi\bigg(R(p,w,s(t))-R(p',w,s'(t))\bigg).
\end{eqnarray}
Therefore, up to an accuracy of order $\mathcal{O}((t/\tau)^{2d+1})$, we obtain
\begin{eqnarray}
&&\big\langle p_{\mathrm{TB}}(m,t|n)\big\rangle_{V}=\sum_{p\in\mathcal{P}(n,m)}\sum_{p'\in\mathcal{P}(m,n)}\int_{\Sigma_t}\mathrm{d}s\int_{\Sigma_t}\mathrm{d}s'\nonumber\\
&&\times\prod_{w\in V}\Phi(R(p,w,s(t)-R(p',w,s'(t))),
\end{eqnarray}
which increases monotonically with $t$.

\subsection{Open CTQW}
The time evolution of a mixed state of a finite dimensional open quantum system in the Markovian regime is described by the Lindblad equation $\dot{\rho}(t)=\mathfrak{L}\rho(t)$, where $\mathfrak{L}$ is given by 
\begin{eqnarray}
&&\mathfrak{L}\rho(t)=-\mathrm{i}[H,\rho(t)]\nonumber\\
&&\qquad\qquad+\sum_{k}\left(L^{\ }_k\rho(t)L^{+}_k-\frac{1}{2}\left\{L_k^{+}L^{\ }_k,\rho(t)\right\}\right),
\label{eq:lindblad}
\end{eqnarray}
wherein the $L_k$'s are linear operators acting on the Hilbert space $\mathcal{H}$ of the system. Choosing a basis $\ket{1},\dots,\ket{d}$ in $\mathcal{H}$, the super-operator $\mathfrak{L}$ becomes a map between $d\times d$ matrices. Choosing the basis $E_{nm}=\ket{n}\bra{m}$ in the space of $d\times d$ matrices, $\mathfrak{L}$ can be represented as a $d^2\times d^2$ matrix with entries $\mathrm{Tr}[E^+_{nm}\mathfrak{L}E_{kl}]$. 

There is a natural way to realize this matrix as a generalized process taking place on a graph $\mathcal{L}$ obtained from the complete, directed graph $\mathcal{K}_{d^2}$ of $d^2$ nodes, whose vertices are labeled by the matrix units $E_{nm}$ and whose edges $E_{kl}\rightarrow E_{nm}$ are deleted when the corresponding matrix entry $\mathrm{Tr}[E^*_{nm}\mathfrak{L}E_{kl})]$ vanishes. Then, splitting the matrix of $\mathfrak{L}$ into diagonal and purely off-diagonal matrices gives rise to a general walk on $\mathcal{L}$ to which Proposition 1 can be applied.

Assume that $H$ in Eq.~\eqref{eq:lindblad} is a Hamiltonian of a tight-binding model corresponding to the graph $\mathcal{G}=(\mathcal{V},\mathcal{E})$. We would like to construct the graph $\mathcal{L}_\omega$ of a Lindbladian $\mathfrak{L}_\omega$ which corresponds to a quantum stochastic walk (QSW) as has been first introduced in \cite{whitfield2010quantum}. QSW keeps the locality structure of the original unitary process by incorporating Lindblad operators of the form $\ket{n}\bra{m}$, whenever the edge $(n,m)$ is contained within the edge set of $\mathcal{G}$. Note that $\mathcal{G}$ can be recognized as a symmetric directed graph, that is $(n,m)\in\mathcal{E}$ if and only if $(m,n)\in\mathcal{E}$. For convenience, to every vertex $v$ of the complete, directed graph $\mathcal{K}_{d}=(\mathcal{V},\mathcal{F})$ of $d$ nodes, we assign the projection $\hat{v}=\ket{v}\bra{v}$ and to every directed edge $e=(n,m)\equiv n\rightarrow m$ in $\mathcal{F}$, we associate the matrix unit $\hat{e}=\ket{m}\bra{n}$. Then, we have the following equations:
\begin{eqnarray}
H&=&\sum_{v\in \mathcal{V}}V_v\hat{v}+\sum_{e\in \mathcal{E}}\hat{e},\nonumber\\
L_e&=&\hat{e}^+,
\end{eqnarray}
so the Lindbladian of the QSW acts on an arbitrary $d\times d$ matrix $X$ as
\begin{eqnarray}
\mathfrak{L}_\omega X&=&-\mathrm{i}\sum_{v\in \mathcal{V}}V_v[\hat{v},X]-\mathrm{i}\sum_{e\in \mathcal{E}}[\hat{e},X]\nonumber\\
&&\qquad+\,\omega\sum_{e\in\mathcal{E}}\left(\hat{e}^+X\hat{e} -\frac{1}{2}\{\hat{e}\hat{e}^+,X\}\right),
\label{eq:Lomega}
\end{eqnarray}
where $\omega$ measures the relative strength of the coherent and the dissipative parts of the dynamics. For any edge $f$ of $\mathcal{K}_d$, the tail and head vertex of $f$ are denoted by $t_f$ and $h_f$, respectively. Let $u\in \mathcal{V}$. Then, a short calculation gives
\begin{eqnarray}
&&\mathfrak{L}_\omega\hat{u}=-\mathrm{i}\sum_{e\in\mathcal{E}}\left(\delta(t_e,u)-\delta(h_e,u)\right)\hat{e}\nonumber\\
&&\qquad\qquad\ -\omega \mathrm{d}_u\hat{u}+\omega\sum_{e\in\mathcal{E}}\delta(t_e,u)\hat{t}_e,
\end{eqnarray}
where $\mathrm{d}_u$ is the degree of the vertex $u$ within $\mathcal{G}$ and $\delta(x,y)$ is just Kronecker's delta. A similar calculation for any edge $f$ of $\mathcal{K}_d$ results in 
\begin{eqnarray}
&&\mathfrak{L}_\omega\hat{f}=\left(-\mathrm{i}(V_{h_f}-V_{t_f})-\omega\frac{\mathrm{d}_{t_f}+\mathrm{d}_{h_f}}{2}\right)\hat{f}\nonumber\\
&&\qquad-\mathrm{i}\left(\sum_{v\sim_\mathcal{G} h_f}\ket{v}\bra{t_f}-\sum_{v\sim_\mathcal{G} t_f}\ket{h_f}\bra{v}\right),
\end{eqnarray}
where $\sim_\mathcal{G}$ refers to adjacency within $\mathcal{G}$. Grouping together the projections $\ket{v}\bra{v}$ and separately the matrix units $\ket{u}\bra{v}$, $u\neq v$, the $d^2\times d^2$ matrix of $\mathfrak{L}_\omega$ admits the following block-matrix form: 
\begin{equation}
\mathfrak{L}_\omega=
\begin{pmatrix}
\mathfrak{L}_{\mathcal{V}\mathcal{V}}(\omega) & \mathfrak{L}_{\mathcal{V}\mathcal{F}} \\
\mathfrak{L}_{\mathcal{F}\mathcal{V}} & \mathfrak{L}_{\mathcal{F}\mathcal{F}}(\omega) \\
\end{pmatrix}.
\end{equation}
Here, $\mathfrak{L}_{\mathcal{V}\mathcal{V}}(\omega)=-\omega L$, where $L$ is the Laplacian of $\mathcal{G}$, which generates CTRW on $\mathcal{G}$ according to Eq.~\eqref{eq:ctrw}. The non-square matrix $\mathfrak{L}_{\mathcal{F}\mathcal{V}}$ is equal to $\mathrm{i}I$, where $I$ is the signed incidence matrix of $\mathcal{G}$ within $\mathcal{K}_d$, that is for a given edge $e\in\mathcal{F}$ and a vertex $v\in \mathcal{V}$,
\begin{equation}
I_{ev}=
\begin{cases}
1 & \text{if}\ h_e=v\ \text{and}\ e\in \mathcal{E}, \\
-1 & \text{if}\ t_e=v\ \text{and}\ e\in \mathcal{E},\\
0 & \text{if}\ e\notin \mathcal{E}.\\
\end{cases}
\end{equation}
Furthermore, we have $\mathfrak{L}_{\mathcal{V}\mathcal{F}}=\mathrm{i}I^+$. Finally, $\mathfrak{L}_{\mathcal{F}\mathcal{F}}(\omega)$ is the sum of the diagonal matrix composed of the entries  
\begin{equation}
V_f(\omega)=-\frac{\omega \mathrm{d}_{h_f}+2\mathrm{i}V_{h_f}}{2}-\frac{\omega \mathrm{d}_{t_f}-2\mathrm{i}V_{t_f}}{2},
\label{eq:vdef}
\end{equation}
and the matrix $-\mathrm{i}\hat{A}$, where $\hat{A}$ is the signed adjacency matrix given by the entries 
\begin{equation}
\hat{A}_{ef}=
\begin{cases}
1 & \text{if}\  t_e=t_f\ \text{and}\ (h_f,h_e)\in\mathcal{E},\\
-1 & \text{if}\ h_e=h_f\ \text{and}\ (t_e,t_f)\in\mathcal{E},\\
0 & \text{otherwise}.\\ 
\end{cases}
\end{equation}
Therefore, the block structure of $\mathfrak{L}_\omega$ is of the form 
\begin{equation}
\mathfrak{L}_\omega=
\begin{pmatrix}
-\omega D+\omega A & \mathrm{i}I^+\\
\mathrm{i}I & V(\omega)-\mathrm{i}\hat{A} \\
\end{pmatrix},
\end{equation}
where $D$ is the degree matrix of $\mathcal{G}$ and $V(\omega)$ is the diagonal matrix defined in Eq.~\eqref{eq:vdef}. This determines $\mathcal{L}_\omega$, the graph of $\mathfrak{L}_\omega$ completely. 

The shortest paths of $\mathcal{L}_\omega$ can be illustrated in the following way. Suppose that we would like to find the shortest directed path connecting vertices of $\mathcal{L}_\omega$ labeled by matrix units $\ket{n}\bra{m}$ and $\ket{k}\bra{l}$. Pick up two copies of the original graph $\mathcal{G}$. Any pair of vertices which formed by vertices of the distinct copies of $\mathcal{G}$ represents a node of $\mathcal{L}_\omega$ and appears as a crosslink between nodes of the copies of $\mathcal{G}$ (see Fig.~\ref{fig:fig5}). Then, to find the shortest directed path connecting $\ket{n}\bra{m}$ to $\ket{k}\bra{l}$ one manipulates the endpoints of the crosslink initially representing $\ket{n}\bra{m}$ by moving its endpoints through neighboring vertices of $\mathcal{G}$ according to the following rules: On the one hand, if $\omega=0$, one is allowed to move only one endpoint of the crosslink in each step. On the other hand, when $\omega>0$, the rules of moving the crosslinks are the same except of those which correspond to projections: the endpoints of the crosslink $\ket{n}\bra{n}$ can be changed within one step to obtain $\ket{m}\bra{m}$ if $m$ is adjacent to $n$ within $\mathcal{G}$. 

The pictorial representation of the the shortest paths of $\mathcal{L}_\omega$ described above gives the following distance of $E_{nm}$ and $E_{kl}$ within $\mathcal{L}_\omega$. When $\omega=0$, then

\begin{equation}
d_{\mathcal{L}_0}\bigl(E_{nm},E_{kl}\bigr)=d_\mathcal{G}(n,k)+d_\mathcal{G}(m,l)
\end{equation}
and the number of such $p$ paths is
\begin{eqnarray}
\hspace*{-5mm} \ell_{\mathcal{L}_0}\bigl(E_{nm},E_{kl}\bigr){=}\ell_\mathcal{G}(n,k)\ell_\mathcal{G}(m,l)\binom{d_{\mathcal{L}_0}\bigl(E_{nm},E_{kl}\bigr)}{d_\mathcal{G}(n,k)}
\end{eqnarray}
all of them carrying the amplitude
\begin{equation}
\Phi_p[\mathfrak{L}_0]=\mathrm{i}^{d_\mathcal{G}(n,k)}(-\mathrm{i})^{d_{\mathcal{G}}(m,l)}.
\end{equation}
However, if $\omega>0$, then 
\begin{eqnarray}
d_{\mathcal{L}_\omega}\bigl(E_{nm},E_{kl}\bigr)&=&\min_{(u,v)\in\mathcal{V}\times\mathcal{V}}\bigl[d_{\mathcal{L}_0}\bigl(E_{nm},E_{uu}\bigr)\nonumber\\
&&\ +d_\mathcal{G}(u,v)+d_{\mathcal{L}_0}\bigl(E_{vv},E_{kl}\bigr)\bigr].
\label{eq:distmin}
\end{eqnarray}
\begin{center}
    \centering
    \begin{figure}[!h]
    \includegraphics[width = 0.42\textwidth]{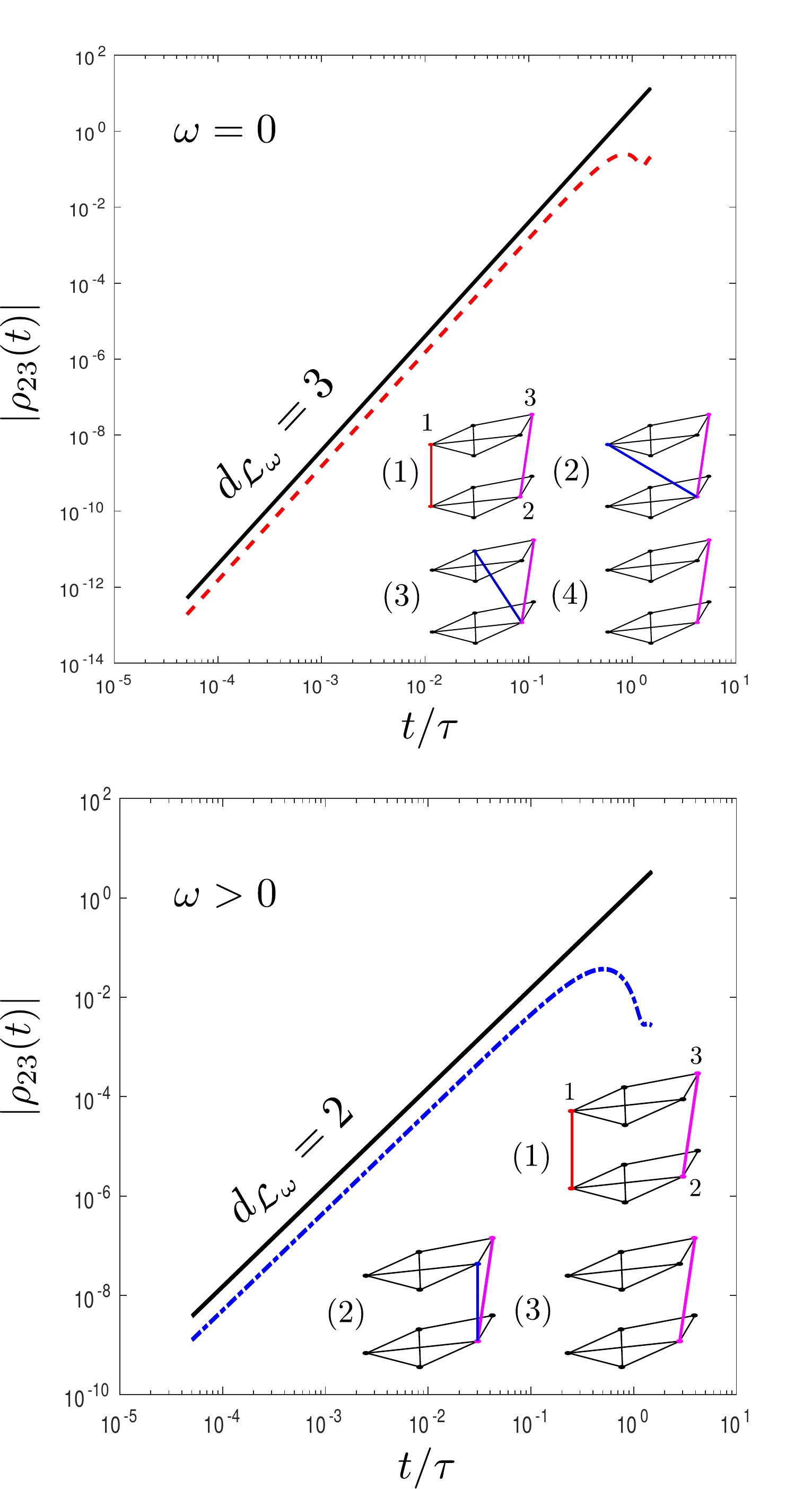}
    \caption{The time evolution of a mixed state according to Eq.~\eqref{eq:Lomega} on a particular small graph. The initial state is $\ket{1}\bra{1}$. It appears as a crosslink between two copies of the graph. Shortest paths $\ket{1}\bra{1}$ $\ket{n}\bra{n}$ initial crosslink and a $\ket{k}\bra{l}$ final crosslink illustrated by magenta lines are detailed step-by-step on the 
            subfigures in the right corners. An instance of the possible series of the intermediate states are illustrated by blue crosslines. It is visible from the $|\rho_{lk}(t)|$ transition probabilities, that in case of $\omega=0$
            the shortest possible path distance allowed by the rules of movements is larger than in case of $\omega>0$. (Color online.)}    
    \label{fig:fig5}
    \end{figure}
\end{center}
Every pair $(u,v)\in\mathcal{V}\times\mathcal{V}$ which minimizes the r.h.s of Eq.~\eqref{eq:distmin} defines a directed path connecting the vertex corresponding to $E_{nm}$ to the vertex corresponding to $E_{kl}$: This path $p$ is a concatenation of three paths $p_1$, $p_2$ and $p_3$ within $\mathcal{L}_\omega$: $p_1$ is a shortest path connecting $E_{nm}$ to $E_{uu}$ within $\mathcal{L}_0$, $p_3$ is a shortest path connecting $E_{vv}$ to $E_{kl}$ within $\mathcal{L}_0$ and finally $p_2$ connects the projections $\ket{u}\bra{u}$ to $\ket{v}\bra{v}$ along projections $\ket{p_{2,1}}\bra{p_{2,1}},\dots,\ket{p_{2,n}}\bra{p_{2,n}}$ for which $(p_{2,1},\dots,p_{2,n})$ is a shortest path connecting $u$ to $v$ within $\mathcal{G}$. The number of the shortest paths with such a pair $(u,v)$ is equal to 
\begin{equation}
\ell_{\mathcal{L}_0}\bigl(E_{nm},E_{uu}\bigr)\ell_{\mathcal{G}}(u,v)\ell_{\mathcal{L}_0}\bigl(E_{vv},E_{kl}\bigr),
\label{eq:elll2}
\end{equation}
and such a $p$ path carries the amplitude
\begin{equation}
\Phi_p[\mathfrak{L_\omega}]=\mathrm{i}^{d_\mathcal{G}(n,u)}(-\mathrm{i})^{d_{\mathcal{G}}(m,u)}\mathrm{i}^{d_\mathcal{G}(v,k)}(-\mathrm{i})^{d_{\mathcal{G}}(v,l)}\omega^{d_\mathcal{G}(u,v)}.
\label{eq:distmin2}
\end{equation}

In the finite dimensional linear space $M_d$ of $d\times d$ complex matrices, the map which assigns $\mathrm{Tr}(A^+B)$ to every pair of matrices $A$ and $B$ is a hermitian scalar product turning $M_d$ to a Hilbert space, the Hilbert-Schmidt space of $d\times d$ matrices. For the sake of brevity, we denote this scalar product by $\braket{A}{B}_{\mathrm{HS}}$. This also induces the norm $\|A\|_{\mathrm{HS}}=\sqrt{\braket{A}{A}_{\mathrm{HS}}}$. Every super-operator $\mathfrak{K}$ acting linearly on $M_d$ obtains a norm similar to that have been introduced in section \ref{sec:proof}:
\begin{equation}
\|\mathfrak{K}\|=\max_{A\neq 0}\frac{\|\mathfrak{K}A\|_{\mathrm{HS}}}{\|A\|_{\mathrm{HS}}},
\end{equation}
and this norm satisfies the usual properties. Therefore, we can apply the methods of section \ref{sec:proof} in order to obtain the short time evolution of density matrix entries. 

If $\mathfrak{L}_\omega$ is the Lindbladian of a QSW, the short time asymptotics of the time evolution of the density matrix entries $\rho_{nm}(t)$ of an initial pure state $\ket{u}\bra{u}$ can be obtained by the approximation of the scalar product $\bra{E_{nm}}\mathrm{e}^{\mathcal{L}_0t}\ket{E_{uu}}_\mathrm{HS}$. If $\omega=0$, we obtain
\begin{eqnarray}
&&\rho_{nm}(t)=\ell_{\mathcal{G}}(n,u)\ell_{\mathcal{G}}(m,u) \binom{d_{\mathcal{L}_0}(E_{nm},E_{uu})}{d_\mathcal{G}(m,u)}\nonumber\\
&&\qquad\qquad\qquad\times\frac{(\mathrm{i}t)^{d_\mathcal{G}(n,u)}(-\mathrm{i}t)^{d_\mathcal{G}(m,u)}}{d_\mathcal{G}(n,u)!d_{\mathcal{G}}(m,u)!}\nonumber\\
&&\qquad\qquad\qquad+\mathcal{O}\left(t^{d_\mathcal{G}(n,u)+d_\mathcal{G}(m,u)+1}\right).
\label{eq:rhoL0}
\end{eqnarray}
Note that, for $n\neq m$, this equation is not the same as the product of the approximative formulae of $\bra{n}U\ket{u}$ and $\bra{u}U^*\ket{m}$ as given by Proposition 1. But this is not surprising if one notes that $\mathfrak{L}_0=-\mathrm{i}[H,\bullet]$ acting on the Hilbert-Schmidt space of $\mathcal{B}(\mathcal{H})$ is different than $H$ acting on $\mathcal{H}$. Not even the timescales where Eq.~\eqref{EQ3} and Eq.~\eqref{eq:rhoL0} are applicable are the same. Indeed, if $\lambda_n$ denote the eigenvalues of $\mathcal{H}$, then $\tau^{-1}_H=\max|\lambda_n|$, while $\tau^{-1}_{\mathfrak{L}_0}=\max|\lambda_n-\lambda_m|$, clearly indicating $\tau_H>\tau_{\mathfrak{L}_0}$ whenever $H$ is non-negative. 
\vspace{10pt}
If $\omega>0$, then the application of Eq.~\eqref{eq:distmin} and Eq.~\eqref{eq:elll2} enables us to write
\begin{eqnarray}
&&\rho_{nm}(t)=\nonumber\\
&&\quad\sum_{(u,v)}\ell_{\mathcal{G}}(u,v)\ell_{\mathcal{L}_0}\bigl(E_{nm},E_{uu}\bigr)\ell_{\mathcal{L}_0}\bigl(E_{vv},E_{uu}\bigr)\nonumber\\
&&\quad\quad\times \dfrac{(\mathrm{it})^{d_\mathcal{G}(n,u)}(-\mathrm{it})^{d_{\mathcal{G}}(m,u)}(\mathrm{it})^{d_\mathcal{G}(v,u)}}{d_{\mathcal{L}_\omega}\bigl(E_{nm},E_{uu}\bigr)!}\omega^{d_\mathcal{G}(u,v)}\nonumber\\
&&\quad\quad + \mathcal{O}\left(t^{d_{\mathcal{L}_\omega}\bigl(E_{nm},E_{uu}\bigr)+1}\right),
\label{eq:rhoLomega}
\end{eqnarray}
where the sum runs over the the pairs $(u,v)\in\mathcal{V}\times\mathcal{V}$ which are the minimizers of the r.h.s of Eq.~\eqref{eq:distmin2}.

Results of comparison of numerical  calculations and approximative formulae Eq.~\eqref{eq:rhoL0} and Eq.~\eqref{eq:rhoLomega} in case of the small graph introduced in section \ref{ssec:chiral} are depicted in Fig.~\ref{fig:fig5}.

\section{Conclusion and Outlook}\label{sec:conclusion}

We studied the short time asymptotics of quantum dynamics on graphs considering both coherent and
open continuous time quantum walks, including time-dependent couplings. 
In the case of non-chiral coherent CTQWs, the short time asymptotics is completely 
determined by the topology of the graph. The transition probabilities follow the short time asymptotics

\begin{equation}
    |\!\bra{x}U(t)\ket{y} \!|^2{=}\left(\frac{\ell(x,y)}{d(x,y)!}\right)^2t^{2d(x,y)}{+}\mathcal{O}(t^{2d(x,y)+1}).
\end{equation}

Furthermore, it has been shown that the on-site potential does not affect this asymptotics. Similar results can be obtained for chiral CTQWs, but it is important to note that introducing time-reversal breaking terms may increase the exponent of the first non-vanishing term in the transition probabilities.  We have also studied open CTQWs through stochastic quantum walks and proved that the short time dynamics of these systems are also significantly altered when they are coupled to the environment. 

Finally, we would like to mention possible future applications of our results. We hope to be able to use these for designing quantum networks with efficient transport properties. In particular, the fact that one can reduce some transition probabilities by tuning the phases of the hopping amplitudes in chiral walks could be utilized to design certain preferred (and non-preferred) transportation directions.  Similar features for designing \mbox{(non-)preferred} directions  or even generating dark states by tuning the hopping were already studied in Refs. \cite{zimboras2013quantum, todtli2016continuous, sett2019};  our methods could provide a more systematic treatment of this. Another possible application of our results comes from the observation that the actual measurement of the short time asymptotics resulting in the distance of the nodes can be interpreted as a distance oracle. 
Such an oracle can be used to reconstruct the graph of the Lindbladian of the system. One may hope that such a reconstruction would be efficient, as it is known that there exist randomized algorithms for the reconstruction problem with query complexity $\mathcal{O}(n^{3/2})$ \cite{distoracles}. 
These two possible directions are left for future work.

\section*{Acknowledgements}

The work was supported by the Hungarian National Research, Development and Innovation Office (NKFIH) through Grants No. K120660, K109577, K124351, K124152, KH129601, and 
the Hungarian Quantum Technology National Excellence Program (Project No. 2017-1.2.1-NKP-2017- 00001). ZZ also acknowledges  support from the J\'anos Bolyai Scholarship of the Hungarian Academy of Sciences.\\

\appendix
\section{Gauge transformation of chiral walks}
\label{app:A}
Let $\mathcal{G}=(V,\mathcal{E})$ be a directed graph without self-loops. Assume that whenever the edge $(u,v)$ with tail $u$ and head $v$ appears in $\mathcal{G}$, then $(v,u)\in \mathcal{E}$ also holds. Let $z_{uv}$ denote the complex phase of modulus one attached to the edge $(u,v)$. Denote by $H$ the hermitian matrix containing entries $H_{uv}=r_{uv}z_{uv}$, where $r_{uv}>0$ if $(u,v)\in \mathcal{E}$ and zero otherwise. By hermiticity, we have $z_{uv}=\overline{z}_{vu}$. Let us denote the matrix composed of the numbers $r_{uv}$ by $R$. We prove the following statement.\\  

\textbf{Proposition}. There exists a unitary, diagonal matrix $\Lambda$ such that $\Lambda^\dagger H\Lambda =R$ if and only if along any closed, directed path $p=(p_0,p_1,\dots,p_n)$, $p_0=p_n$ the product of complex phases $\phi_p$ is equal to one 
\begin{equation}
\phi_{p}=z_{p_0p_1}\cdots z_{p_{n-1}p_n}=1.
\label{cond}
\end{equation}
\par
\textbf{Proof}. Assume that $\Lambda^\dagger H\Lambda =R$ holds and let $\Lambda=\mathrm{diag}(\lambda_1,\dots,\lambda_{|V|})$. Then, $z_{uv}=\overline{\lambda}_u\lambda_v$, so for a given closed path $p=(p_0,p_1,\dots,p_n)$ we have
\begin{eqnarray}
\phi_p=z_{p_0p_1}\cdots z_{p_{n-1}p_n}&=&\overline{\lambda}_{p_0}\lambda_{p_1}\cdot\overline{\lambda}_{p_1}\lambda_{p_2}\cdots \overline{\lambda}_{p_{n-1}}\lambda_{p_n}\nonumber\\
&=&\overline{\lambda}_{p_0}\lambda_{p_n}=1.
\end{eqnarray}
In the reversed direction of the statement, assume that the condition holds. Choose a vertex $\star$ and for each other vertex $u$, a path $p^{(u)}=(\star,p^{(u)}_1,\dots,p^{(u)}_{n_u})$ connecting $\star$ to $u=p^{(u)}_{n_u}$.  Let $\Lambda$ defined through the diagonal entries $\lambda_{\star}=1$ and $\lambda_{u}=\phi_{p^{(u)}}$. Then, if $u\neq v$,
\begin{eqnarray}
(\Lambda^\dagger H\Lambda)_{uv}&=&\overline{\lambda}_{u}R_{uv}\lambda_{v}=\overline{\phi}_{p^{(u)}}z_{uv}\phi_{p^{(v)}}r_{uv}\nonumber\\
&=&r_{nm}\overline{\phi}_{q},
\end{eqnarray}
where $q$ is the closed path 
\begin{eqnarray}
q=(\star,p^{(u)}_{1},\dots,p^{(u)}_{n_{m-1}},u,v,p^{(v)}_{n_v-1},\dots,p^{(v)}_{1},\star).
\end{eqnarray}
Since the condition of Eq.~\eqref{cond} holds, we have $\phi_q=1$, thus the statement is proved.  $\blacksquare$ \par

Note that such a global trivialization of $U(1)$ phases can be always achieved for Hamiltonians corresponding to tree graphs, 
since the walks generated by $\Lambda^\dagger H\Lambda$ and $H$  have identical site-to-site transition probabilities \cite{zimboras2013quantum}, a chiral walk on a tree has  identical short time asymptotics as its non-chiral counterpart.

\bibliography{asymptotics}
\bibliographystyle{apsrev4-1}

\end{document}